\documentclass[lettersize,journal]{IEEEtran}
\usepackage{amsmath,amsfonts}
\usepackage{algorithmic}
\usepackage{algorithm}
\usepackage{array}
\usepackage[caption=false,font=normalsize,labelfont=sf,textfont=sf]{subfig}
\usepackage{textcomp}
\usepackage{stfloats}
\usepackage{url}
\usepackage{verbatim}
\usepackage{graphicx}
\usepackage{cite}
\usepackage{multirow}
\usepackage{tabularx,booktabs}
\usepackage{xcolor,colortbl}
\usepackage{mathrsfs}
\definecolor{firstcolor}{rgb}{0.95,0.69,0.51}
\definecolor{secondcolor}{rgb}{0.99,0.9,0.6}
\definecolor{thirdcolor}{rgb}{0.99,0.90,0.60}
\newcommand{\cf}[1]{\cellcolor{firstcolor} #1}
\newcommand{\cs}[1]{\cellcolor{secondcolor} #1}
\newcommand{\uli}[1]{\underline{#1}}

\newcommand{\TableRQours}{
\begin{table*}[!t]
\caption{Per-scene image quality comparison on the RTR dataset.}
\vspace{-3mm}

\begin{tabularx}{\textwidth}{l|l|>{\centering\arraybackslash}X>{\centering\arraybackslash}X>{\centering\arraybackslash}X>{\centering\arraybackslash}X>{\centering\arraybackslash}X>{\centering\arraybackslash}X|>{\centering\arraybackslash}X}
\toprule
Metric & Method & Loft       & Gundam     & Porcelain  & Ship       & Bookcase1  & Bookcase2  & Average \\
\hline
\multirow{8}{*}{PSNR $\uparrow$}   

& Ref-NeRF      & 33.39      & 28.48      & 31.58      & 29.68      & 29.76      & 26.50      & 29.90   \\
& MS-NeRF       & 33.42      & 29.20      & 31.68      & 30.72      & 29.71      & 24.94      & 29.94   \\
& NeRFRen       & 32.29      & 28.66      & 32.19      & 29.67      & 26.28      & 25.97      & 29.18   \\
& 3DGS          & 34.10      & 28.84      & 31.16      & 31.16      & 30.24      & 25.45      & 30.16   \\
& Spec-Gaussian & 34.13      & 29.30      & \cs{32.27} & \cf{31.20} & \cs{30.73} & \cs{26.45} & \cs{30.68}  \\
& EnvGS         & \cs{34.25} & \cf{29.69} & 31.99      & 31.11      & 29.63      & 25.77      & 30.41 \\
& Ours          & \cf{34.63} & \cs{29.64} & \cf{32.52} & \cs{31.14} & \cf{30.83} & \cf{28.51} & \cf{31.21}   \\
\cline{2-9} 
& 3DGS-MCMC     & 35.09      & 30.60      & \uli{33.68}& 31.75      & 31.76      & 27.48      & 31.73   \\
& Ours-MCMC     & \uli{35.15}& \uli{30.85}& 33.58      & \uli{32.12}& \uli{32.21}& \uli{28.71}& \uli{32.10} \\
\hline
\multirow{8}{*}{SSIM $\uparrow$}   

& Ref-NeRF      & 0.937      & 0.912      & 0.917      & 0.916      & 0.913      & 0.859      & 0.909   \\
& MS-NeRF       & 0.957      & 0.928      & 0.940      & 0.940      & 0.940      & 0.894      & 0.933   \\
& NeRFRen       & 0.951      & 0.923      & 0.934      & 0.919      & 0.885      & 0.847      & 0.910   \\
& 3DGS          & 0.965      & 0.935      & 0.953      & 0.953      & \cs{0.949} & \cs{0.917} & 0.945   \\
& Spec-Gaussian & \cs{0.966} & \cs{0.938} & \cs{0.966} & \cs{0.953} & 0.948      & 0.915      & \cs{0.948}   \\
& EnvGS         & 0.962      & 0.937      & 0.959      & 0.948      & 0.942      & 0.910      & 0.943 \\
& Ours          & \cf{0.967} & \cf{0.940} & \cf{0.967} & \cf{0.954} & \cf{0.951} & \cf{0.928} & \cf{0.951}   \\
\cline{2-9} 
& 3DGS-MCMC     & \uli{0.968}& 0.950      & 0.967      & 0.960      & 0.957      & 0.927      & 0.955   \\
& Ours-MCMC     & 0.967      & \uli{0.952}& \uli{0.968}& \uli{0.962}& \uli{0.958}& \uli{0.931}& \uli{0.956}   \\
\hline

\multirow{8}{*}{LPIPS $\downarrow$}

& Ref-NeRF      & 0.262      & 0.256      & 0.254      & 0.252      & 0.262      & 0.288      & 0.262   \\
& MS-NeRF       & 0.189      & 0.212      & 0.192      & 0.184      & 0.176      & 0.203      & 0.193   \\
& NeRFRen       & 0.209      & 0.229      & 0.216      & 0.242      & 0.287      & 0.313      & 0.250   \\
& 3DGS          & 0.165      & 0.188      & 0.153      & 0.150      & 0.147      & \cs{0.144} & 0.158   \\
& Spec-Gaussian & \cf{0.158} & \cs{0.183} & \cf{0.106} & \cf{0.142} & \cf{0.144} & 0.146      & \cs{0.147}   \\
& EnvGS         & 0.168      & 0.191      & 0.136      & 0.152      & 0.186      & 0.173      & 0.168 \\
& Ours          & \cs{0.164} & \cf{0.178} & \cs{0.112} & \cs{0.146} & \cs{0.146} & \cf{0.134} & \cf{0.146}   \\
\cline{2-9} 
& 3DGS-MCMC     & 0.167      & 0.175      & 0.121      & 0.137      & 0.137      & 0.137      & 0.146   \\
& Ours-MCMC     & \uli{0.163}& \uli{0.174}& \uli{0.115}& \uli{0.135}& \uli{0.135}& \uli{0.136}& \uli{0.143}   \\
\bottomrule

\end{tabularx}
\label{table:rqours}
\end{table*}
}

\newcommand{\TableMirror}{
\begin{table}[h]

\caption{Image quality comparison on the Mirror-NeRF dataset.}
\vspace{-3mm}

\begin{tabularx}{\columnwidth}{l|>{\centering\arraybackslash}X>{\centering\arraybackslash}X>{\centering\arraybackslash}X}
\toprule

              & PSNR $\uparrow$  & SSIM $\uparrow$  & LPIPS $\downarrow$ \\ 
\hline
Ref-NeRF      & 25.12      & 0.752      & 0.376 \\
MS-NeRF       & 26.58      & 0.830      & 0.261 \\
Mirror-NeRF   & 26.25      & 0.840      & \cs{0.187} \\
NeRFReN       & 25.29      & 0.77       & 0.337  \\
3DGS          & 26.82      & 0.876      & 0.196 \\
Spec-Gaussian & \cs{26.89} & \cs{0.867} & 0.198 \\
EnvGS         & 26.45      & 0.866      & 0.191  \\
Ours          & \cf{28.01} & \cf{0.889} & \cf{0.170} \\
\hline
3DGS-MCMC     & 27.57      & 0.882      & 0.188 \\
Ours-MCMC     & \uli{29.10}& \uli{0.904}& \uli{0.155} \\

\bottomrule
\end{tabularx}
\label{table:rqmirrornerf}
\end{table}
}

\newcommand{\TableSpeed}{
\begin{table}[t]

\caption{Training time, rendering speed (FPS), and number of Gaussians on the RTR dataset.}
\vspace{-3mm}

\begin{tabularx}{\columnwidth}{l|>{\centering\arraybackslash}X>{\centering\arraybackslash}X>{\centering\arraybackslash}X}
\toprule

Method        & Training Time$\downarrow$ & FPS$\uparrow$  & Gaussian Number$\downarrow$ \\ 
\hline
Ref-NeRF      & 6h       & 0.04     & / \\
MS-NeRF       & 3h       & 0.06     & / \\
NeRFReN       & 50h      & 0.04     & / \\
3DGS          & \cs{21m} & \cf{338} & 659k \\
Spec-Gaussian & \cf{16m} & 181      & \cs{568k} \\
EnvGS         & 2h       & 13      & 1076k  \\
Ours          & 33m      & \cs{225} & \cf{560k} \\

\bottomrule
\end{tabularx}
\label{table:speed}
\end{table}
}

\newcommand{\TableAblationRegulation}{
\begin{table}[t]

\caption{Ablation study: quantitative impact of each component on image quality}
\vspace{-3mm}

\begin{tabularx}{\columnwidth}{l|>{\centering\arraybackslash}X>{\centering\arraybackslash}X>{\centering\arraybackslash}X}
\toprule
Ablations    & PSNR$\uparrow$ & SSIM$\uparrow$  & LPIPS$\downarrow$  \\ 
\hline
w/o multi-stage optimization                & 28.51    & 0.930    & 0.154  \\
w/o $\mathcal{L}_d$ and $\mathcal{L}_c$     & 29.33    & 0.938    & 0.144  \\
w/o $\mathcal{L}_d$                         & 29.53    & 0.937    & 0.145  \\
w/o $\mathcal{L}_c$                         & 29.56    & 0.938    & 0.143  \\
w/o opacity perturbation                    & 29.59    & 0.938    & 0.144  \\
w/o reflection intensity                    & 29.46    & 0.936    & 0.146  \\
w/o Fresnel-based reflection                 & 29.59    & 0.935    & 0.143  \\
Full model                                  & \cf{29.83}    & \cf{0.941}    & \cf{0.140}  \\

\bottomrule
\end{tabularx}
\label{table:ablation}
\end{table}
}

\newcommand{\FigRqOurs}{
\begin{figure*}[tbp]
  \begin{center}
    \includegraphics[width=0.99\textwidth]{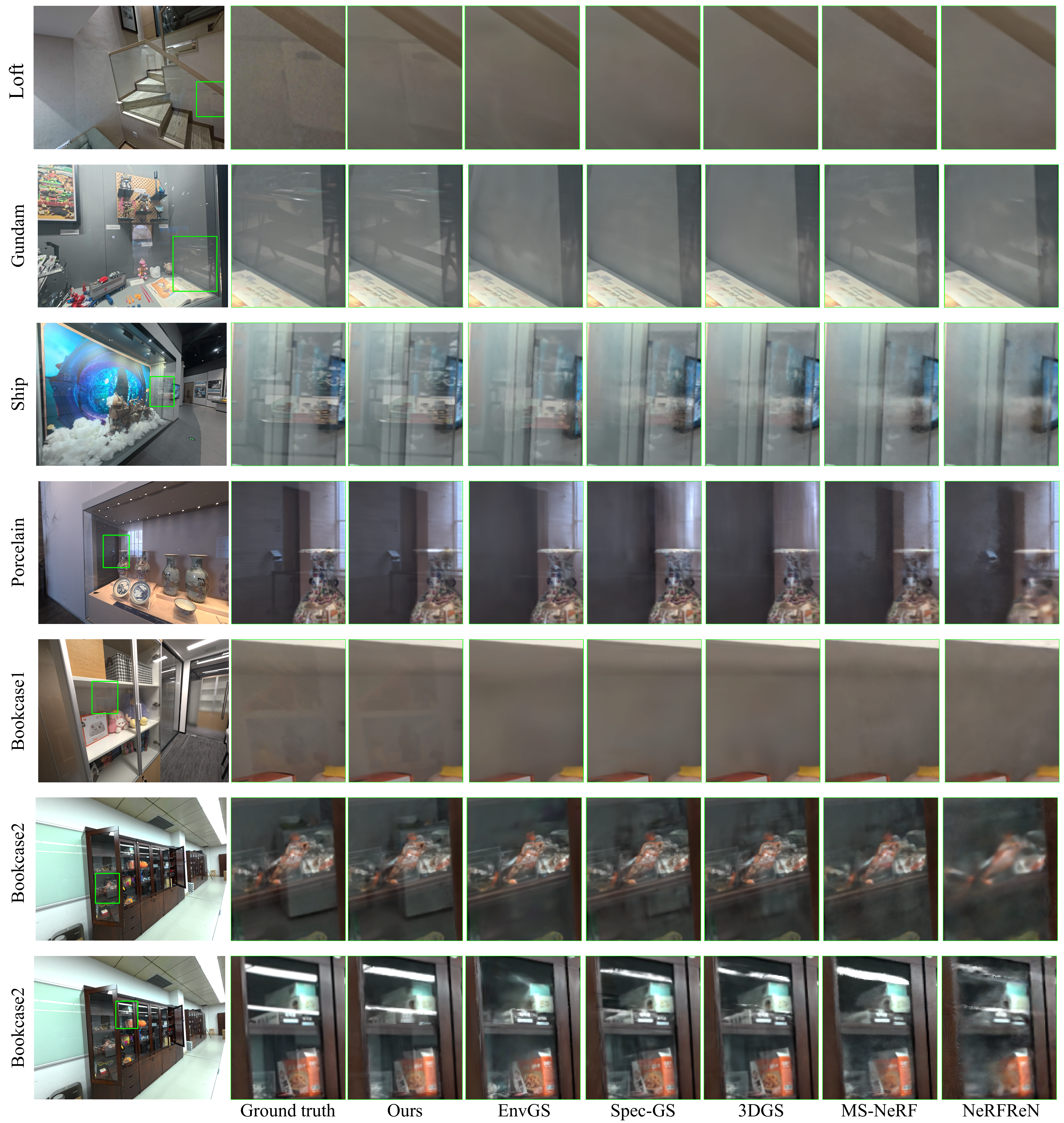}
  \end{center}
  \vspace{-5mm}
  \caption{Qualitative comparison with EnvGS~\cite{xie2024envgs}, Spec-Gaussian~\cite{spec-gaussian}, 3DGS~\cite{3dgs}, MS-NeRF~\cite{ms-nerf} and NeRFReN~\cite{nerfren} on the RTR dataset. We show images rendered from novel test viewpoints. Compared to these baselines, our method produces more accurate and detailed reflections, and yields results closest to the ground truth. }
  \label{fig:rqours}
\end{figure*}
}

\newcommand{\FigRqmirrornerf}{
\begin{figure*}[!t]
  \begin{center}
    \includegraphics[width=0.99\textwidth]{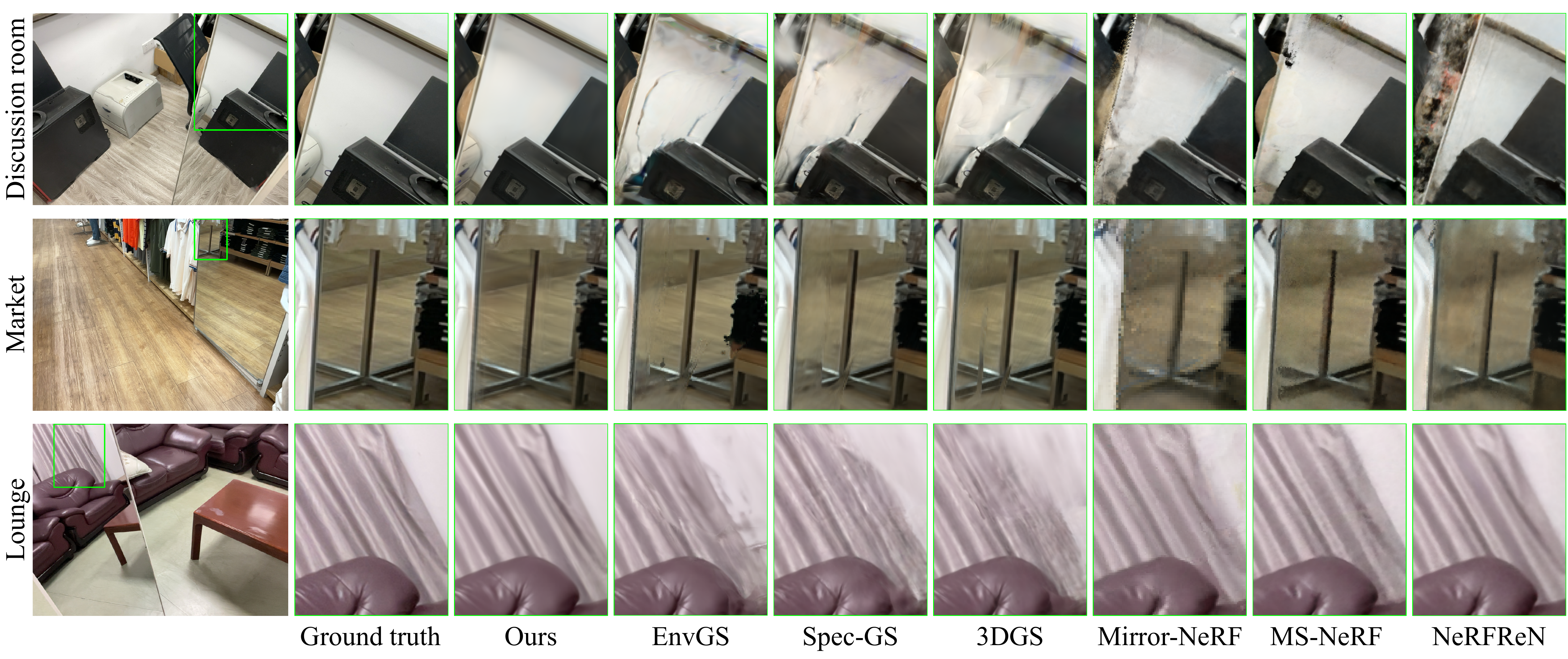}
  \end{center}
  \vspace{-5mm}
  \caption{Qualitative comparison on Mirror-NeRF dataset. Regions with noticeable differences are cropped and shown side-by-side for clear comparison.}
  \label{fig:rqmirror}
\end{figure*}
}

\newcommand{\FigDeours}{
\begin{figure*}[t]
    \centering
    \includegraphics[width=0.99\textwidth]{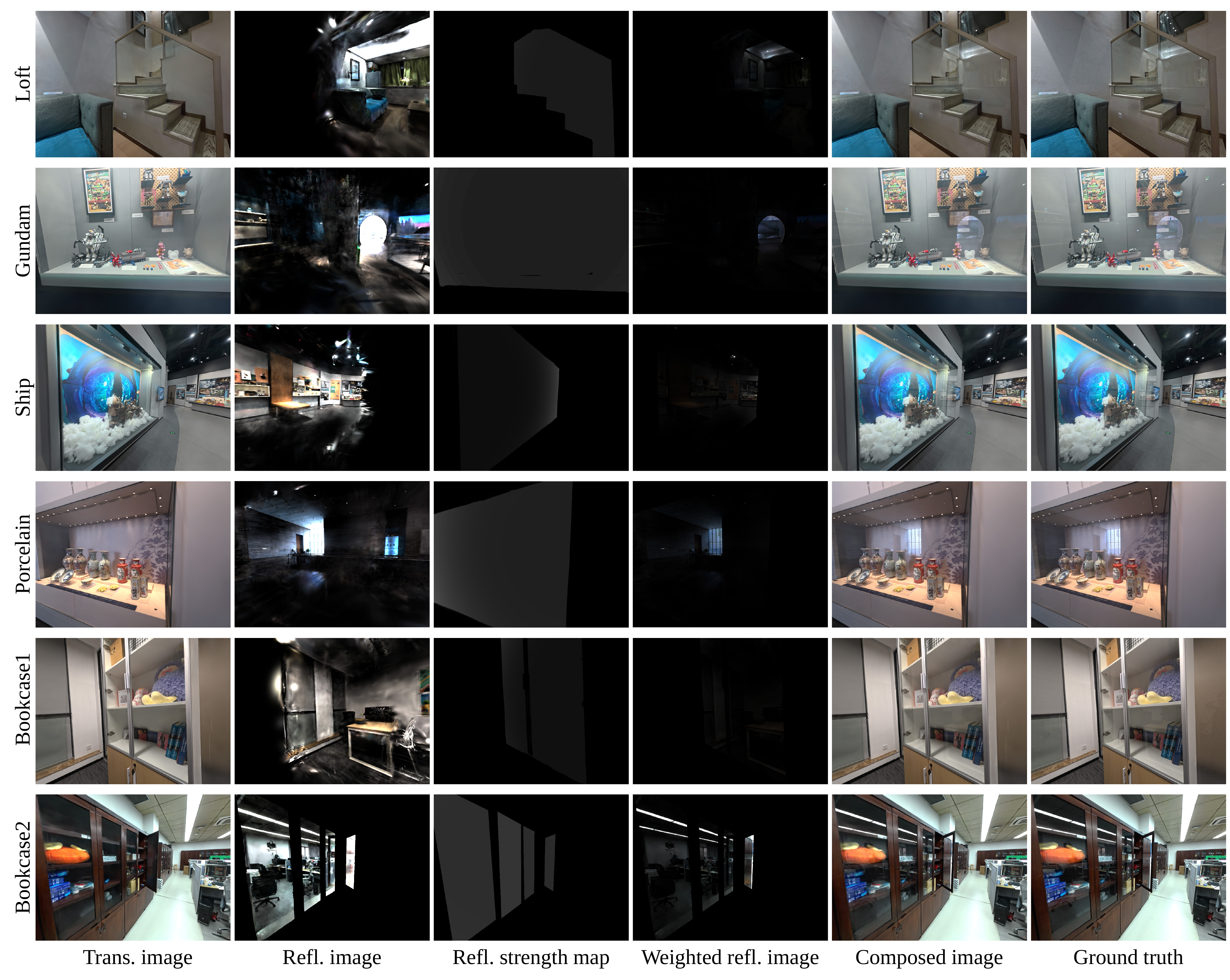}
  \caption{Decomposition results on RTR dataset. From left to right, there are transmission image $\mathbf{C}_r$, reflection image $\mathbf{C}_r$, reflection strength map $\mathbf{R}$, weighted reflection image $\mathbf{R}\cdot\mathbf{C_r}$ and final image $\mathbf{C}$. Our method can achieve natural transmission-reflection decomposition and produce high-quality novel view synthesis results.}
  \label{fig:decomours}
\end{figure*}
}

\newcommand{\FigDeccompare}{
\begin{figure}[h]
  \begin{center}
    \includegraphics[width=0.49\textwidth]{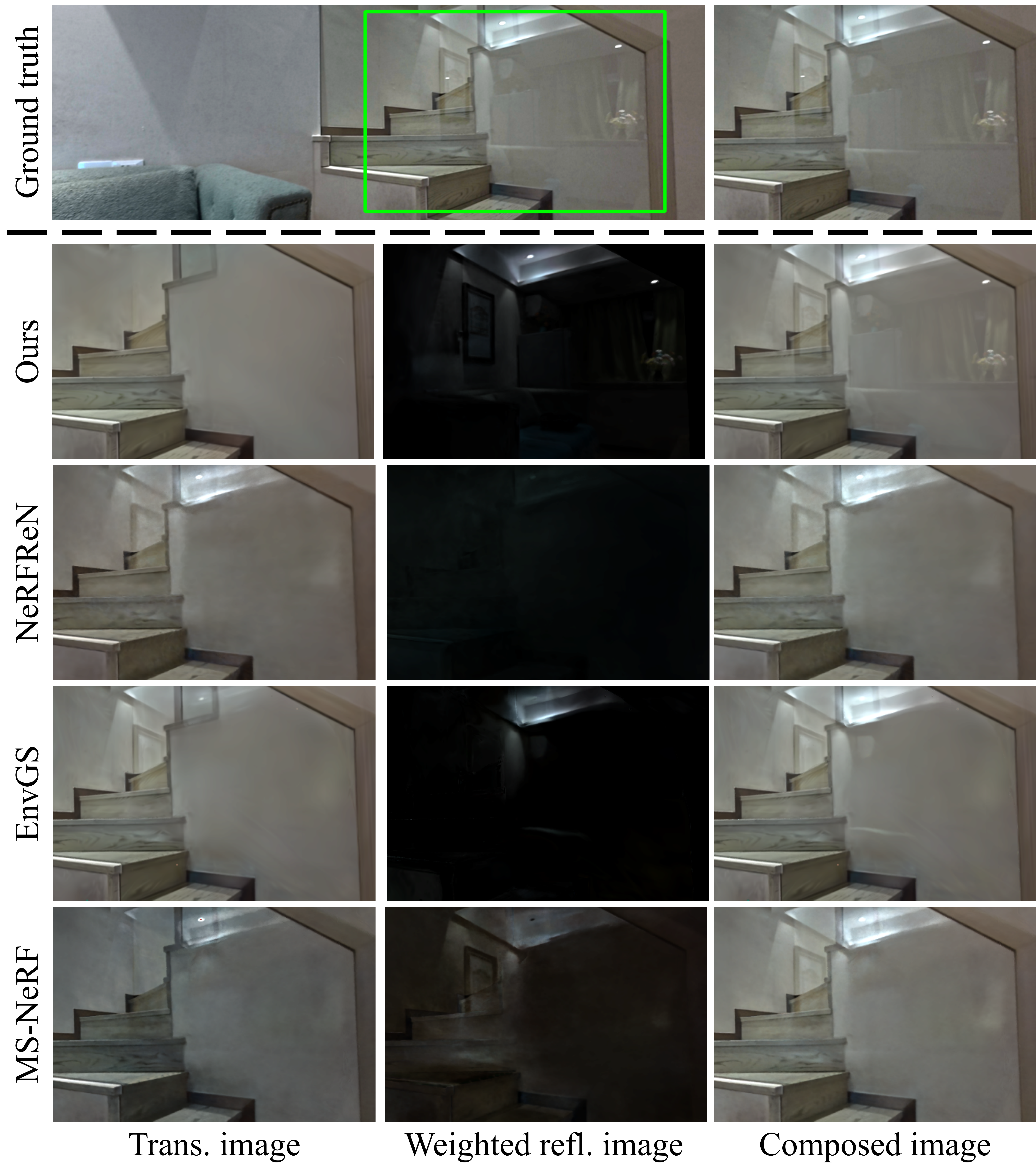}
  \end{center}
  \vspace{-5mm}
  \caption{Decomposition results in \textit{Loft} compared with and NeRFReN~\cite{nerfren}, MS-NeRF~\cite{ms-nerf}, and EnvGS~\cite{xie2024envgs}. Our method achieves comparable decomposition results and better novel view synthesis quality.}
  \label{fig:deccompare}
\end{figure}
}

\newcommand{\FigAbReflInsensity}{
\begin{figure}[t]
  \begin{center}
    \includegraphics[width=0.49\textwidth]{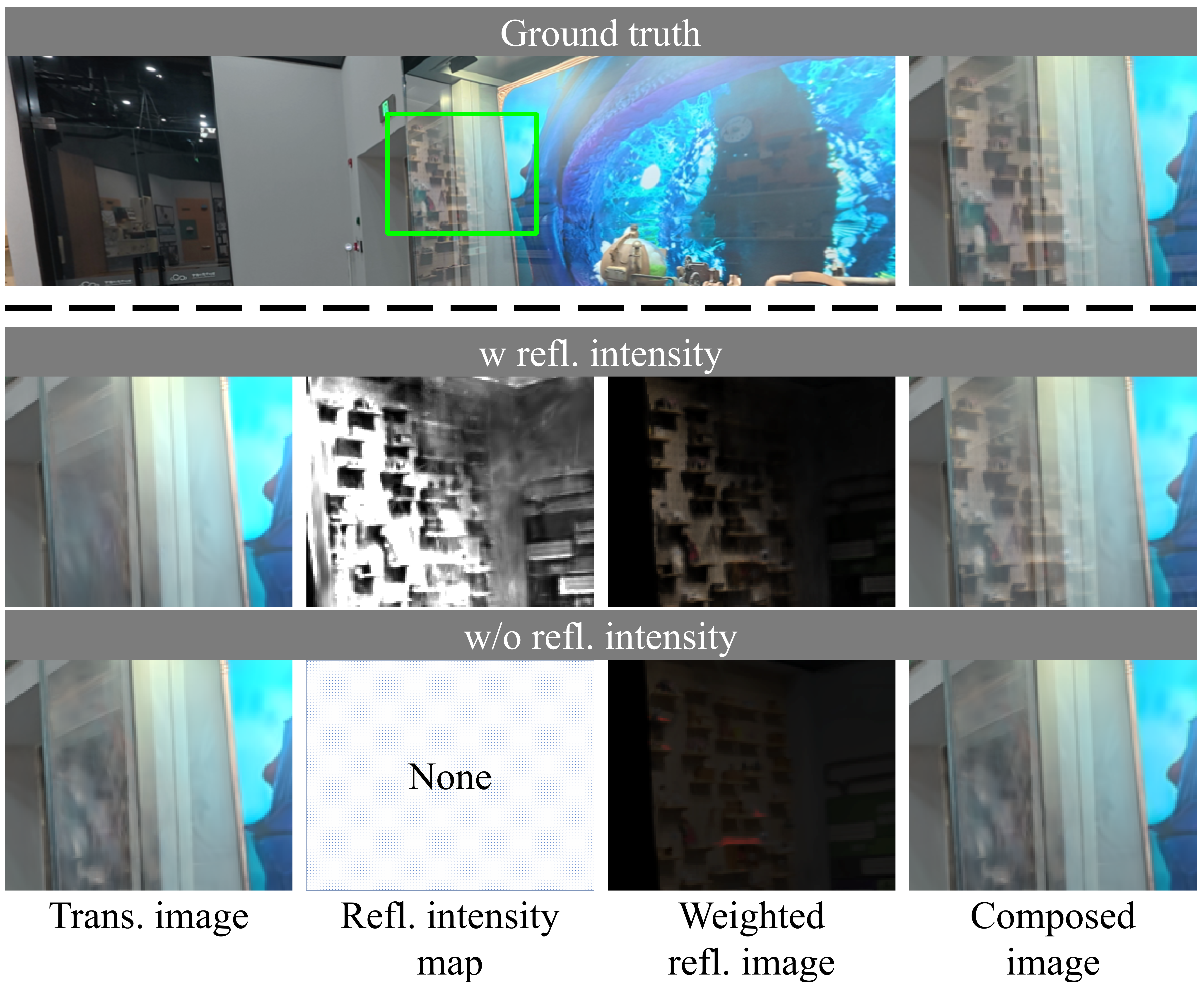}
  \end{center}
  \vspace{-5mm}
  \caption{Without the reflection intensity, the reflection color input to our model no longer adheres to the required linear color space, which hampers accurate reflection modeling and leads to degraded rendeirng quality.}
  \label{fig:ablationreflection_intensity}
\end{figure}
}

\newcommand{\FigAbRegulation}{
\begin{figure}[!t]
  \begin{center}
    \includegraphics[width=0.49\textwidth]{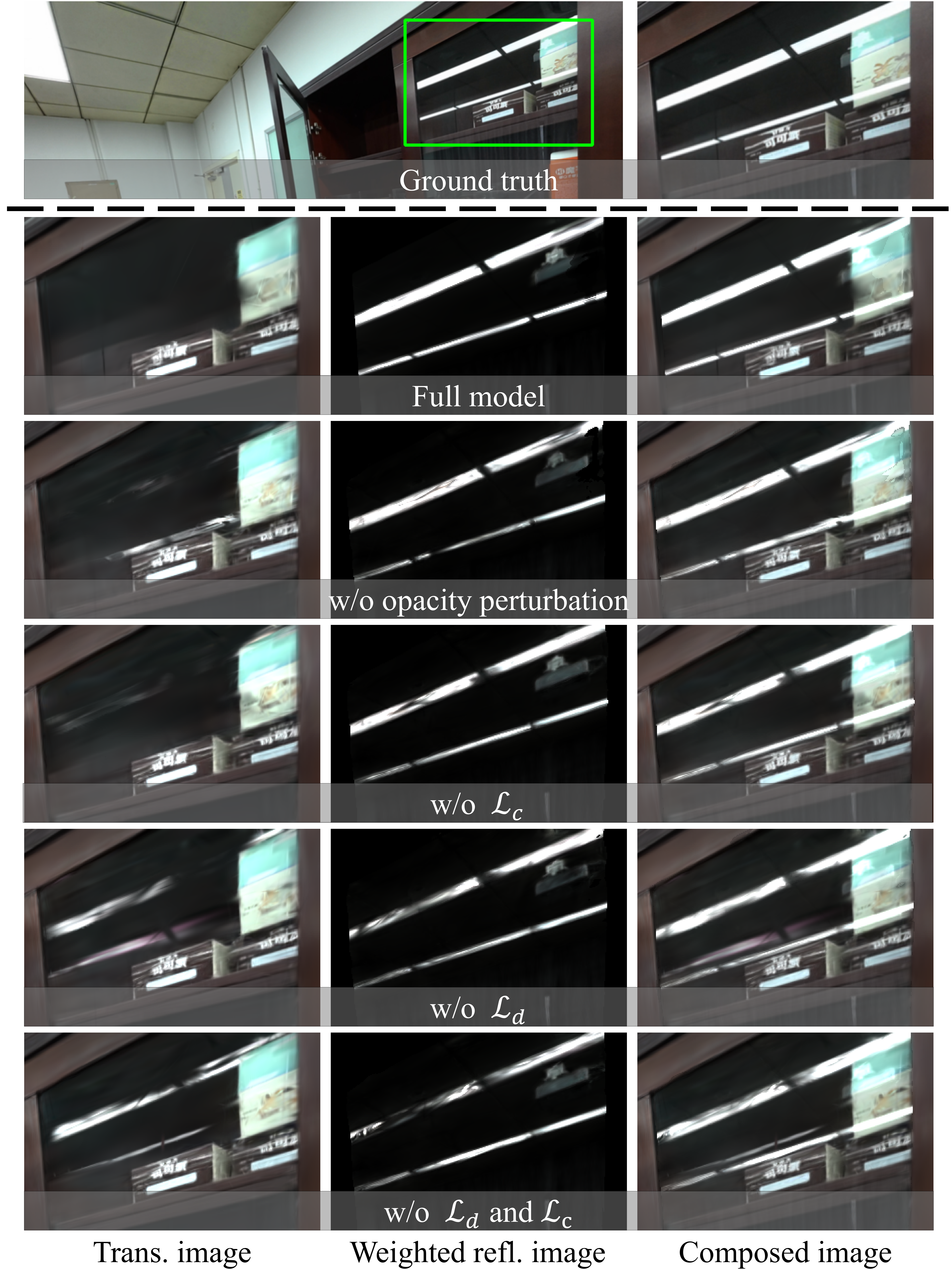}
  \end{center}
  \vspace{-5mm}
  \caption{Ablation results of regulations and opacity perturbation in \textit{Bookcase2}. The removal of either component leads to degraded decomposition, highlighting the importance of their joint contribution to realistic rendering.}
  \label{fig:ablationregulation}
\end{figure}
}

\newcommand{\FigPipeline}{
\begin{figure*}[tbp]
  \begin{center}
    \includegraphics[width=0.99\textwidth]{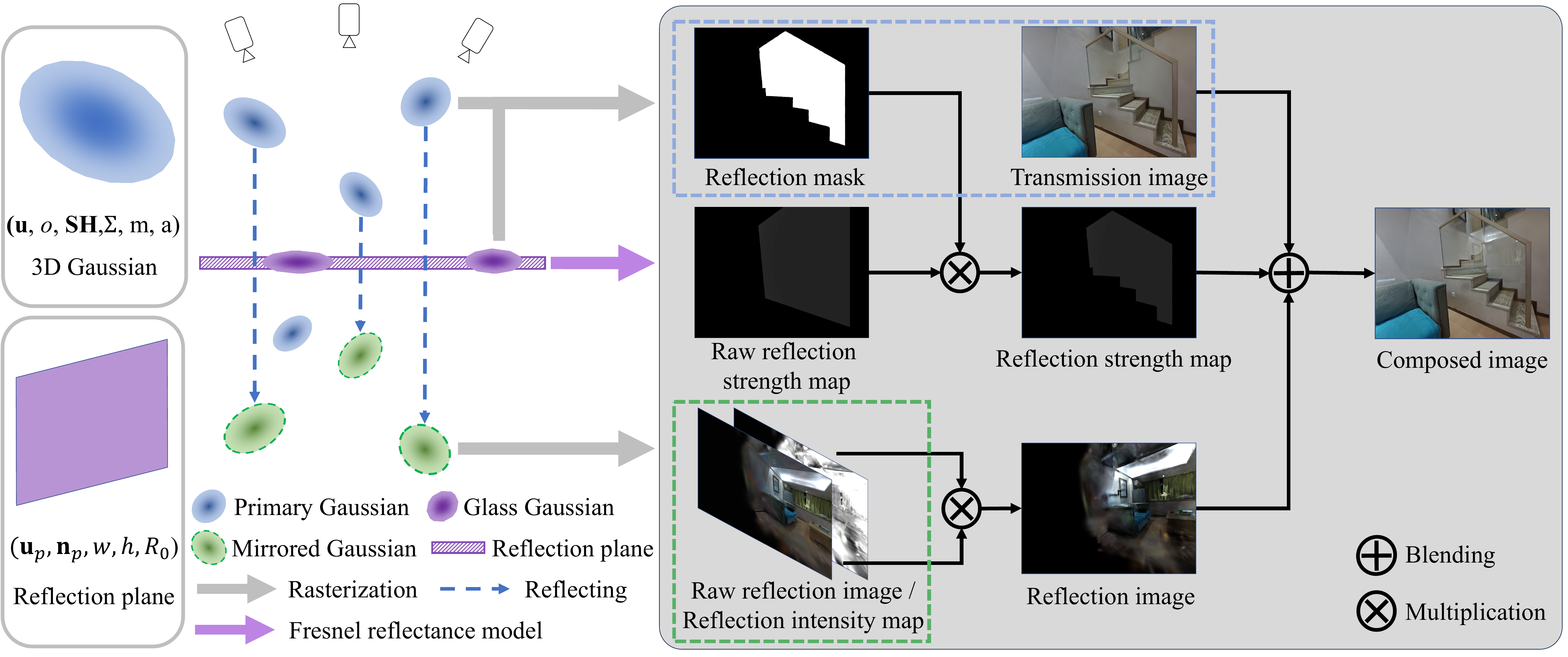}
  \end{center}
  \vspace{-5mm}
  \caption{Our rendering pipeline. The pipeline consists of two rendering passes. In the first pass, we render the transmission image $\mathbf{C}_t$ using primary Gaussians, and generate the reflection mask $\mathbf{M}$ by rendering both primary and glass Gaussians. We then compute the raw reflection strength map $\mathbf{R}_{raw}$ using Fresnel reflectance model, and multiply it with the reflection mask $\mathbf{M}$ to obtain the reflection strength map $\mathbf{R}$. In the second pass, we mirror the primary Gaussians across the reflection plane $\mathbf{P}$ to produce the mirrored Gaussians, which are rasterized to generate the raw reflection image $\mathbf{C}_{raw}$ and the reflection intensity map $\mathbf{A}$. These two are multiplied to produce the reflection image $\mathbf{C}_r$. Finally, we blend the transmission image $\mathbf{C}_t$ and the reflection image $\mathbf{C}_r$ using the reflection strength map $\mathbf{R}$ to produce the final image. }
  \label{fig:pipeline}
\end{figure*}
}

\newcommand{\FigAbMultiStage}{
\begin{figure}[t]
  \begin{center}
    \includegraphics[width=0.49\textwidth]{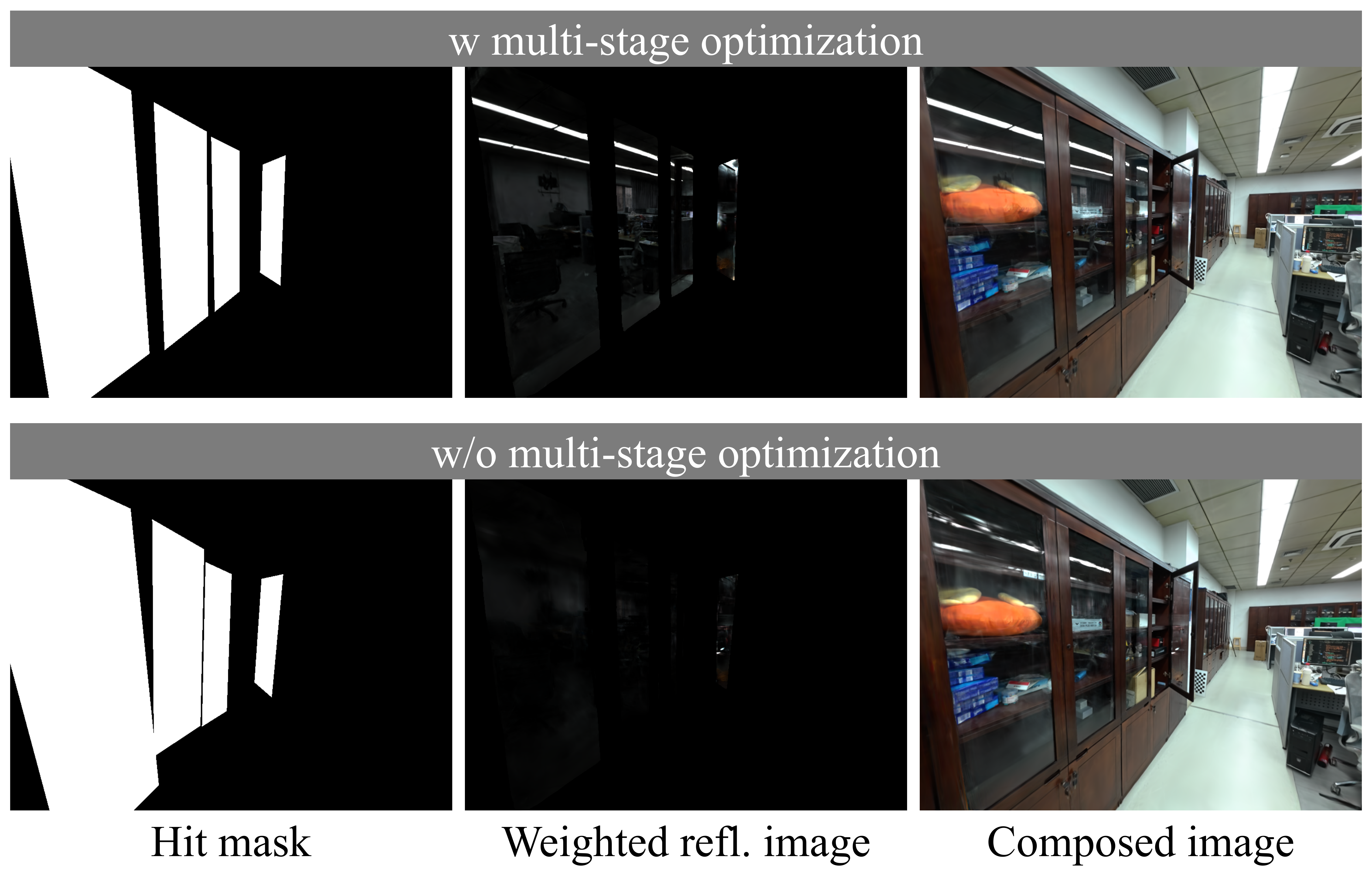}
  \end{center}
  \vspace{-5mm}
  \caption{Disabling multi-stage optimization leads to reflection plane drift and poor separation of reflection and transmission, demonstrating that the it is crucial for accurate reflection modeling and high-quality rendering.}
  \label{fig:ablationstage}
\end{figure}
}

\newcommand{\FigGSFloater}{
\begin{figure}[h]
  \begin{center}
    \includegraphics[width=0.49\textwidth]{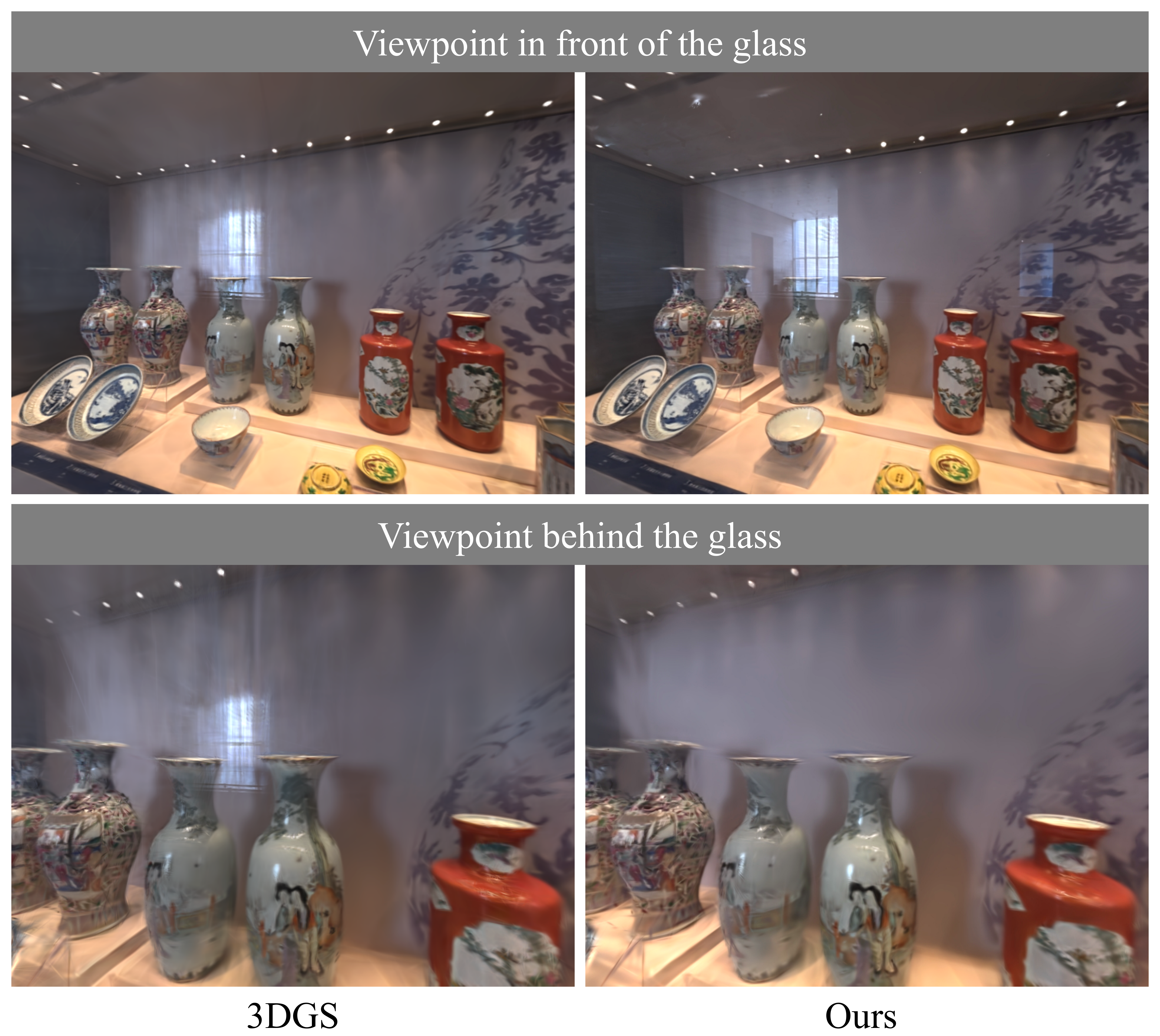}
  \end{center}
  \vspace{-5mm}
  \caption{Rendering comparison of 3DGS (left column) and our method (right column) from two viewpoints — in front of the glass (top row) and behind the glass (bottom row). Our method effectively disentangles reflection and transmission components: clear reflections occur solely on the glass’s front side; no reflections are visible from behind. By comparison, 3DGS fails to accurately model reflections on transparent glass, producing many floaters of ``reflection ghost'' far from the real surface.}
  \label{fig:gsfloater}
\end{figure}
}

\newcommand{\FigAbSchlick}{
\begin{figure}[t]
  \begin{center}
    \includegraphics[width=0.49\textwidth]{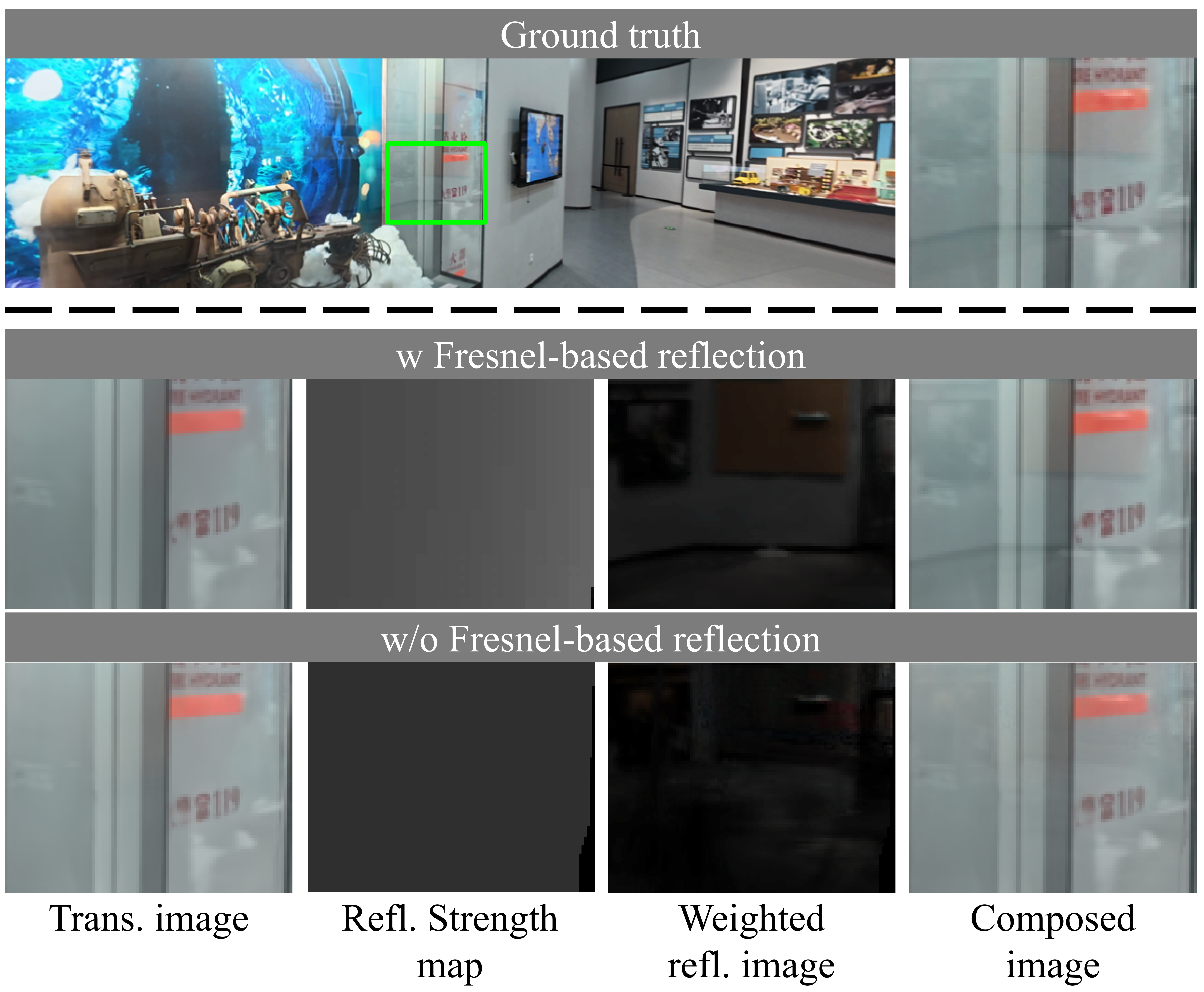}
  \end{center}
  \vspace{-5mm}
  \caption{Ablation results of Fresnel-based reflection in \textit{Ship}. Without it, the model fails to capture view-dependent reflection strength, especially at grazing angles, which results in blurrier outputs.}
  \label{fig:ablationschlick}
\end{figure}
}

\begin{document}

\title{TR-Gaussians: High-fidelity Real-time Rendering of Planar Transmission and Reflection with 3D Gaussian Splatting}

\author{Yong~Liu, Keyang~Ye, Tianjia~Shao, and Kun~Zhou,~\IEEEmembership{Fellow,~IEEE}



\thanks{Yong Liu, Keyang Ye, Tianjia Shao, Kun Zhou are with the State Key Lab of CAD \& CG, Zhejiang University, Hangzhou 310058, China.}
\thanks{E-mail: \{yongliu6, yekeyang, tjshao\}@zju.edu.cn, kunzhou@acm.org.}
\thanks{Manuscript received April 19, 2021; revised August 16, 2021.}}

\markboth{Journal of \LaTeX\ Class Files,~Vol.~14, No.~8, August~2021}%
{Shell \MakeLowercase{\textit{et al.}}: A Sample Article Using IEEEtran.cls for IEEE Journals}


\maketitle


\begin{abstract}
We propose Transmission-Reflection Gaussians (TR-Gaussians), a novel 3D-Gaussian-based representation for high-fidelity rendering of planar transmission and reflection, which are ubiquitous in indoor scenes. 
Our method combines 3D Gaussians with learnable reflection planes that explicitly model the glass planes with view-dependent reflectance strengths. Real scenes and transmission components are modeled by 3D Gaussians and the reflection components are modeled by the mirrored Gaussians with respect to the reflection plane. The transmission and reflection components are blended according to a Fresnel-based, view-dependent weighting scheme, allowing for faithful synthesis of complex appearance effects under varying viewpoints.
To effectively optimize TR-Gaussians, we develop a multi-stage optimization framework incorporating color and geometry constraints and an opacity perturbation mechanism. Experiments on different datasets demonstrate that TR-Gaussians achieve real-time, high-fidelity novel view synthesis in scenes with planar transmission and reflection, and outperform state-of-the-art approaches both quantitatively and qualitatively.

\end{abstract}




\section{Introduction}
\label{sec:introduction}





\IEEEPARstart{T}{ransparent} glass panes (e.g., windows and showcases) are ubiquitous in indoor scenes in people's daily life, exhibiting complex combinations of transmission and reflection at their surfaces. 
It is of critical importance to accurately model such optical phenomena in photorealistic novel view synthesis (NVS) of indoor scenes.
While there has been a rapid development of NVS in recent years, where 3D Gaussian Splatting (3DGS)~\cite{3dgs} has emerged as the state-of-the-art method, demonstrating high rendering quality in many types of scenes, it still struggles to faithfully reconstruct indoor scenes with glass panes.
The fundamental reason is, given the training images with both transmission and reflection effects, 3DGS simply overfits them using low-opacity Gaussians, without correctly distinguishing or modeling the two components separately. As a result, while training views may be fitted well, under test views the reflection often exhibits noisy artifacts (e.g., \textit{Bookcase2} in Fig.~\ref{fig:rqours}) and can be completely missing in regions unobserved by training views (e.g., \textit{Loft} in Fig.~\ref{fig:rqours}).

To the best of our knowledge, there has not been 3DGS-related works studying the high-fidelity novel-view rendering of transparent glass panes in indoor scenes. The most relevant work is MirrorGaussian~\cite{mirrorgaussian}, which focuses only on mirrors with pure reflection without considering transmission. Recently several NeRF-based methods were proposed aiming to address the mixed phenomenon of transmission and reflection. They decompose the refection and transmission either by regarding the scene as a single shell without multiple surfaces~\cite{nerfren}, or by suppressing gradient similarities between primary and reflected colors~\cite{gao2024planar}, which only holds on limited views or lacking robustness on textureless regions. Moreover, these methods inevitably suffer from NeRF's inherent computational inefficiency.




\IEEEpubidadjcol

In this paper, we propose a novel representation of 3D Gaussians named Transmission-Reflection Gaussians (TR-Gaussians)
enabling high-fidelity modeling of light transmission and reflection on transparent glass panes, which can be rendered in real time during novel view synthesis. TR-Gaussians consist of a set of primary Gaussians representing the real scene, a reflection plane representing the glass pane and modeling the view-dependent reflection strengths, and a set of glass Gaussians marking the reflective regions on the glass pane. We apply the Fresnel reflectance model on the reflection plane, where the reflection plane is associated with the learnable properties of position, orientation and base reflectance, and the reflection strength on each point of the plane is computed as the ratio of reflected light intensity relative to the incident light intensity based on these properties.
The rendering of TR-Gaussians involves two passes. In the first pass, the primary Gaussians are used to render the real scene and the transmission image, and the primary and glass Gaussians are utilized together to render a reflection mask on the reflection plane marking the arbitrarily shaped reflective regions. We also use the reflection plane to compute the reflection strength map and filter it with the reflection mask. In the second pass, by
leveraging the symmetry of planar reflections, we generate the mirrored Gaussians by reflecting the primary Gaussians about the reflection plane, and the mirrored Gaussians are rasterized to obtain the reflection image. 
The final image is obtained by blending the transmission image and reflection image based on the reflection strength map.

We design effective optimization strategies to optimize TR-Gaussians. First, to restrain the primary Gaussians from wrongly fitting the reflective regions with low-opacity Gaussians, which are commonly floating far from the real surfaces (see Fig.~\ref{fig:gsfloater} for example), we introduce a depth variance loss to enforce the primary Gaussians to be distributed close to the surface, by minimizing the distances between the primary Gaussians' center depths and the primary Gaussian rendered depths. 
By combining the depth variance loss with the image loss~\cite{3dgs} and the gradient conflict loss~\cite{gao2024planar}, it is supposed that we can correctly decompose the reflection regions and real scenes. 
However, we find that relying solely on loss functions for decomposition may still get stuck in local optima. 
To this end, we propose a multi-stage optimization framework. In the first stage, all Gaussians are optimized together as vanilla 3DGS and then the reflection planes and glass Gaussians are initialized. In the second stage we start the joint optimization of the reflection planes along with the primary and glass Gaussians, but ignore the gradients from the reflection images to primary Gaussians so as to avoid the interference from reflection colors, because we find incorporating reflection colors together will yield wrongly placed reflection planes, which will cause the reflection cannot be correctly modeled and is degenerated to vanilla 3DGS. In the final stage, the primary Gaussians, glass Gaussians and reflection planes are jointly optimized to obtain the final image.
Furthermore, we propose an opacity perturbation strategy to periodically add noise to the opacity of primary Gaussians during optimization. With this simple operation, we can simultaneously perturb both the depth variance and gradient similarity, thereby escaping from local optima and progressively refining the decomposition. 

Extensive experiments on both our captured real dataset and public mirror reflection dataset demonstrate that TR-Gaussians can achieve high-fidelity novel view rendering results of indoor scenes with complex transmission and reflection, and outperforms state-of-the-art methods both quantitatively and qualitatively.
In summary, our contributions include:
\begin{itemize}
    \item We present a novel 3D Gaussian representation to model high-fidelity transmission and reflection from glass panes. 
    \item We design effective optimization strategies to achieve high-quality decomposition of the reflection regions and real scenes.
    \item We provide a real-world indoor scene dataset containing common transparent glass panes with pronounced transmission-reflection effects. 
\end{itemize}

\section{Related works}

\subsection{Novel View Synthesis}
Given a set of calibrated images of a scene, the goal of novel view synthesis is to generate high-quality images from new viewpoints~\cite{gortler2023lumigraph,levoy2023light}.
NeRF~\cite{mildenhall2021nerf} has become a significant technique in this field. Leveraging the continuity of MLPs and the volumetric rendering equation, NeRF can achieve end-to-end scene reconstruction and high-quality rendering in novel viewpoints. However, due to the massive MLP evaluation for every ray sample, it suffers from slow training and rendering speed. Subsequent works~\cite{chen2022tensorf,chen2023neurbf,fridovich2022plenoxels,liu2020neural,muller2022instant,sun2022direct,xu2022point} have attempted to employ more advanced representations to tackle these problems. 

Notably, a representative work 3DGS~\cite{3dgs} represents the scene as a collection of explicit 3D Gaussian points, assigning each point opacity and spherical harmonics (SH) coefficients. With the aid of a differentiable rasterizer, 3DGS significantly enhances the training speed and enables high-quality real-time rendering at high resolution. Subsequently, there have been a lot of works focusing on improving the 
rendering quality of 3DGS~\cite{scaffoldgs,yu2024mip,fan2024lightgaussian,3dgs-mcmc,yan2024multi,papantonakis2024reducing}. Mip-Splatting~\cite{yu2024mip} introduces a 3D smoothing filter and a 2D Mip filter to eliminate the aliasing artifacts. 3DGS-MCMC~\cite{3dgs-mcmc} rewrites the densification and pruning strategy as a deterministic state transition of MCMC samples, which effectively eliminates a lot of floaters and improves the rendering quality. These methods focus on improving quality in general scenes but do not specifically address reflection from glass. In this work, our method focuses on reconstructing the transmission and reflection from transparent glass panes, which remains a challenge for previous works. 

\subsection{Reflection Reconstruction}

Reflection modeling has been extensively studied recently, with existing approaches falling broadly into two categories: physically-based modeling and representational enhancement. These approaches have been explored within both NeRF~\cite{boss2021nerd, zhang2022differentiable, zhang2021nerfactor, ref-nerf, ma2024specnerf, mirror_nerf, verbin2024nerf, holland2024tram, gao2024planar, qiu2023looking, nerfren, ms-nerf} and 3DGS~\cite{jiang2024gaussianshader, spec-gaussian, xie2024envgs, moenne20243d, mirrorgaussian,ye20243d} frameworks.

Physically-based methods aim to approximate light transport governed by reflection laws. A common strategy involves using environment maps under the assumption of far-field illumination. Representative works~\cite{boss2021nerd, zhang2021nerfactor, zhang2022differentiable, jiang2024gaussianshader} jointly estimate geometry, material properties, and environment lighting using differentiable rendering techniques. For instance, GaussianShader~\cite{jiang2024gaussianshader} augments each Gaussian with physically meaningful attributes (e.g., diffuse reflectance, tint) and learns an environment map for efficient reflection lookup. 3DGS-DR~\cite{ye20243d} employs deferred rendering techniques to achieve accurate surface normal and a detailed environment map, which further improves the quality of reflection. However, these approaches struggle to model near-field reflections, which violate the far-field assumption, especially in indoor scenes.

To address the limitations of far-field assumptions, another line of physically-based methods simulates reflection by explicitly tracing reflected rays within the scene~\cite{mirror_nerf, verbin2024nerf, holland2024tram, gao2024planar, xie2024envgs, moenne20243d, qiu2023looking} or introducing auxiliary radiance fields to account for reflected content~\cite{nerfren, ms-nerf}. These ray-based techniques can handle near-field and spatially-varying reflections with higher fidelity, but often suffer from high computational cost due to the ray queries.

However, in special cases such as planar mirror surfaces, strong geometric priors allow for more efficient solutions. 
By exploiting the symmetry of planar reflections, MirrorGaussian~\cite{mirrorgaussian} avoids full ray tracing while still preserving high-quality reflections. Specifically, it adopts a dual-rendering scheme that renders both the original scene and its mirror-transformed counterpart, achieving photorealistic mirror effects without expensive ray sampling.
Nevertheless, its formulation is limited to ideal specular mirrors and cannot be generalized to complex transmission-reflection mixtures commonly found in transparent glass.

Complementary to physically-based approaches, another line of work focuses on enhancing the expressiveness of the reflection representation itself. Instead of modeling the underlying light transport, these methods directly regress reflection appearance using MLPs~\cite{ref-nerf, spec-gaussian, ma2024specnerf}. While effective in representing soft highlights and anisotropic effects, the inherent smoothness of MLPs limits their ability to reproduce sharp or spatially discontinuous reflections.

Our method builds on the physically-based reflection model and adopts the planar reflection approximation that supports scenes with both planar mirror and glass-like materials. By unifying reflection and transmission modeling under a single 3DGS framework, our approach enables real-time rendering of complex transparent surfaces while maintaining photorealistic quality.

\subsection{Transmission-Reflection Decomposition}

The problem of separating transmission and reflection components has been studied extensively, both in 2D image processing and in 3D scene reconstruction. Existing methods can be broadly categorized into 2D image-based and 3D scene-based decomposition.

\paragraph{Image-based Decomposition}
Traditional approaches attempt to recover the transmission and reflection layers from one or more images by exploiting various priors, such as gradient sparsity~\cite{levin2007user}, depth cues~\cite{wan2016depth}, or gradient consistency~\cite{li2013exploiting}. For example, Xue \textit{et al.}~\cite{xue2015computational} utilize parallax differences between the two layers from multi-view inputs to aid the decomposition. More recent methods adopt deep learning models to perform single-image separation~\cite{fan2017generic, wan2018crrn, zhang2018single, wei2019single, liu2020learning}, where networks are trained in a supervised or weakly-supervised manner. Zhang \textit{et al.}~\cite{zhang2018single} propose a convolutional network using perceptual losses to remove reflection from a single image. However, how to extend image-based methods to 3D space remains underexplored.

\paragraph{3D Scene-based Decomposition}
With the advent of neural scene representations, several works have attempted to jointly model transmission and reflection in 3D space. NeRFReN~\cite{nerfren} introduces two separate neural radiance fields for transmitted and reflected components, relying on geometric priors to enable disentanglement. However, their approach is restricted to forward-facing scenes with narrow viewpoint coverage, and fails to generalize to more unconstrained environments. MS-NeRF~\cite{ms-nerf} decomposes the scene into multiple parallel feature subspaces without explicitly modeling physical properties, often leading to results that diverge from human perceptual expectations. Gao \textit{et al.}~\cite{gao2024planar} propose to suppress reflection using edge-aware regularization based on color gradient dissimilarity, but this method tends to fail in textureless regions due to the lack of structural cues.

Our method builds upon this line of work by introducing both geometric and photometric constraints for robust decomposition. Specifically, we incorporate a depth variance loss and a gradient conflict loss, which together enforce consistency in 3D space and separation in 2D image gradients. Furthermore, we introduce an \textit{opacity perturbation} operation to escape local minima during optimization, improving robustness and convergence. 
\section{Method}
\FigPipeline

In this section, we first introduce how to combine 3DGS with the reflection plane to model transmission and reflection on glass panes (Section~\ref{sec:dualrendering}). 
We then introduce the regulations for the decomposition (Section~\ref{sec:regulations}).
Finally, we outline the details of our training process (Section~\ref{sec:training}).

\subsection{Transmission and Reflection Modeling}
\label{sec:dualrendering}
We use 3D Gaussians and reflection planes as our scene representation. As shown in Fig.~\ref{fig:pipeline}, to synthesize novel views, the transmission image $\mathbf{C}_t$ rendered by primary Gaussians $\mathcal{G}_i$ and reflection image $\mathbf{C}_r$ rendered by mirrored Gaussians $\hat{\mathcal{G}}_i$ are blended by the the reflection strength map $\mathbf{R}$. Combining these components, we can get the final image $\mathbf{C}$ as follows:
\begin{equation}
    \label{equ:rendering_equ}
    \mathbf{C} = (1 - \mathbf{R}) \cdot \mathbf{C}_t + \mathbf{R} \cdot \mathbf{C}_r.
\end{equation}

We will explain the details of each component.

\subsubsection{Transmission image}
Following the setting of the vanilla 3DGS~\cite{3dgs}, we model the real scene and render the transmission image with the primary Gaussians $\mathcal{G}_i$. Each Gaussian has position $\mathbf{u}_i$, opacity $o_i$, SH coefficients for view-dependent color $\mathbf{c}_i(\mathbf{d})$, 3D covariance $\mathbf{\Sigma}_i = \mathbf{R}_i\mathbf{S}_i\mathbf{S}_i^{\top}\mathbf{R}_i^{\top}$, where $\mathbf{S}_i$ is the scaling matrix and $\mathbf{R}_i$ is the rotation matrix.

During the rendering process, Gaussians are projected to screen space to evaluate Gaussian values $G_i$ with $u_i^{\prime} \in \mathbb{R}^{2}$  and $\sum^{\prime}\in\mathbb{R}^{2\times2}$ via EWA Splatting~\cite{zwicker2001ewa}: 

\begin{equation}
    G_i(u) = exp(-\frac{(\mathbf{u} - \mathbf{u_i^{\prime}})^{\top}\mathbf{\Sigma}_i^{\prime -1}(\mathbf{u}-\mathbf{u}_i^{\prime})}{2}),
\end{equation}
where $\mathbf{u}_i^{\prime}$ and $\mathbf{\Sigma}_i^{\prime}$ are the projected position and covariance matrix in screen space.

Then Gaussians are sorted and alpha blended into pixels to generate the transmission image $\mathbf{C}_t$:
\begin{equation}
    \label{eqn:ct}
    \mathbf{C}_t(\mathbf{u})=\sum_{i=1}\mathbf{c}_i(\mathbf{d})o_iG_i(\mathbf{u})T_i,
    \: T_i=\prod_{j=1}^{i-1}(1-o_jG_j(\mathbf{u})).
\end{equation}

We can also compute the depth map $\mathbf{D}_t$ as:
\begin{equation}
    \mathbf{D}_t(\mathbf{u})=\sum_{i=1}z_io_iG_i(\mathbf{u})T_i / \mathbf{T}_t(u),
    \: \mathbf{T}_t(\mathbf{u})=\sum_{i=1}o_iG_i(\mathbf{u})T_i.
\label{equ:depth_map}
\end{equation}
where $z_i$ denotes the depth of the Gaussian center.

\subsubsection{Reflection strength map}
\label{sec:reflection_strength}

To achieve realistic blending between transmission and reflection, we introduce a \textit{reflection strength map} $\mathbf{R}(\mathbf{u})$, which modulates the contribution of the reflection image in the final composition. This map is defined as the element-wise product of two components: a view-dependent \textit{raw reflection strength map} $\mathbf{R}_{\text{raw}}(\mathbf{u})$ computed through the reflection plane, and a \textit{reflection mask} $\mathbf{M}(\mathbf{u})$ used to refine reflective regions:
\begin{equation}
    \mathbf{R}(\mathbf{u}) = \mathbf{R}_{\text{raw}}(\mathbf{u}) \cdot \mathbf{M}(\mathbf{u}).
\end{equation}

To obtain $\mathbf{R}_{\text{raw}}(\mathbf{u})$, we define a learnable rectangle \textit{reflection plane} to model the transparent glass pane. This plane is parameterized as $(\mathbf{u}_p, \mathbf{n}_p, w, h, R_0)$, where $\mathbf{u}_p$ is the center of the plane, $\mathbf{n}_p$ is the normal vector, $(w, h)$ denote the size, and $R_0$ is the base reflectance coefficient in Schlick's model, which provides an efficient formulation of the Fresnel reflection model. Given the camera viewpoint, we emit primary rays and compute their intersections with the reflection plane, generating a binary hit mask $\mathbf{H}(\mathbf{u})$. For pixels where an intersection exists ($\mathbf{H}(\mathbf{u})=1$), the raw reflection strength is computed as:
\begin{equation}
    \mathbf{R}_{\text{raw}}(\mathbf{u}) = R_0 + (1 - R_0)(1 - \mathbf{n}_p \cdot \mathbf{d})^5,
\end{equation}
where $\mathbf{d}$ denotes the view direction. For pixels with no intersection, $\mathbf{R}_{\text{raw}}$ is set to zero.

However, real-world reflective surfaces such as mirrors or glass panes are rarely perfect rectangle planes and the occlusions of foreground also needs to be considered. To handle this, we add a learnable per-Gaussian property $m_i$ to mark the glass or mirror regions. The reflection mask $\mathbf{M}(\mathbf{u})$ is then rendered in a similar way as the transmission image:
\begin{equation}
    \mathbf{M}(\mathbf{u})=\sum_{i=1}m_io_iG_i(\mathbf{u})T_i. 
\end{equation}

We apply a mask regularization loss $\mathcal{L}_m$ (detailed in Section~\ref{sec:regulations}) to make $\mathbf{M}(\mathbf{u})=1$ in mirrors or glass regions and $\mathbf{M}(\mathbf{u})=0$ in other regions. We define Gaussians with $m_i>0$ as \textit{glass Gaussians}, which are only used in the rendering of $\mathbf{M}(\mathbf{u})$ to indicate glass regions and do not participate in the rendering of transmission or reflection images. 

While the formulation above assumes a single reflection plane for simplicity, our method naturally generalizes to $N \geq 2$ reflection planes through iterative computation of $\mathbf{R}(\mathbf{u})$ and multi-pass rendering. Details of the multi-plane setup are provided in the supplementary material.

\subsubsection{Reflection image}
\label{sec:reflected_image}

To render the reflection image, we need to reflect the primary Gaussians across the reflection plane to generate mirrored Gaussians, by keeping opacity and scale unchanged, and modifying position, rotation and view-dependent color. 
Note that we only reflect the primary Gaussians that are in front of the reflection plane.

First, the position $\hat{\mathbf{u}}_i$ of mirrored Gaussian $\hat{\mathcal{G}_i}$ is computed as:
\begin{equation}
    \mathbf{\hat{\mathbf{u_i}}} = 
    \mathbf{u}_i - 2\frac{(\mathbf{u}_i-\mathbf{u}_p)^{\top}\mathbf{n}_p}
    {\|\mathbf{n}_p\|_{2}}\mathbf{n}_p.
\end{equation}

Next, we reflect $\mathbf{R}_i$ to get the mirrored rotation matrix $\hat{\mathbf{R}}_i$. $\mathbf{R}_i$ can be represented as $[\mathbf{R}_i^1, \mathbf{R}_i^2, \mathbf{R}_i^3]$, where $\mathbf{R}_i^1$, $\mathbf{R}_i^2$, and $\mathbf{R}_i^3$ are the three principal axes of the 3D Gaussian. Therefore, symmetrizing the rotation matrix is equivalent to symmetrizing the three principal axes. And we invert $\hat{\mathbf{R}}_i^1$ to ensure that the three principal axes after reflection still remain in a right-handed coordinate system:
\begin{equation}
    \hat{\mathbf{R}}_i = [-\hat{\mathbf{R}}_i^1, \hat{\mathbf{R}}_i^2, \hat{\mathbf{R}}_i^3].
\end{equation}
\begin{equation}
    \hat{\mathbf{R}}_i^j = 
    \mathbf{R}_i^j - 2\frac{\mathbf{R}_i^{j\top}\mathbf{n}_p}{\|\mathbf{n}_p\|_{2}} \mathbf{n}_p.
\end{equation}

Finally, the view-dependent color $\mathbf{c}_i(\mathbf{d})$ should be adjusted accordingly. The most straightforward method is to reflect the SH coefficients across the reflection plane and then compute the color from the current viewpoint. However, 
mirroring SH coefficients is time-consuming. We reflect the current viewpoint across the reflection plane and compute the color $\hat{\mathbf{c}}_i(\mathbf{d})$ using the reflected view direction:
\begin{equation}
    \hat{\mathbf{c}}_i(\mathbf{d}) = \mathbf{c}_i(\mathbf{\hat{\mathbf{d}}}), 
    \hat{\mathbf{d}} = \mathbf{d} - 2\frac{\mathbf{d}^{\top}\mathbf{n}_p}{\|\mathbf{n}_p\|_{2}} \mathbf{n}_p.
\end{equation}

Then the raw reflection image is computed as follows:
\begin{equation}
    \label{eqn:cr}
    \mathbf{C}_{raw}(\mathbf{u})=\sum_{i=1}\hat{\mathbf{c}}_i(\mathbf{d})\hat{o}_i\hat{G}_i(\mathbf{u})\hat{T}_i.
\end{equation}


Since the Fresnel-based reflection model requires colors to be in linear space, we follow~\cite{gao2024planar} and introduce a reflection intensity parameter $a_i\in\mathbb{R}$ for each Gaussian to compensate for the non-linearity caused by HDR tone mapping. Finally, the reflection image $\mathbf{C}_r(\mathbf{u})$ is computed as:
\begin{equation}
    \mathbf{C}_r(\mathbf{u}) = \mathbf{C}_{raw}(\mathbf{u})\mathbf{A}(\mathbf{u}),
    \: \mathbf{A}(\mathbf{u})=\sum_{i=1}\hat{a}_i\hat{o}_i\hat{G}_i(\mathbf{u})\hat{T}_i.
\end{equation}

\subsection{Regulation}
\label{sec:regulations}

\subsubsection{Depth variance loss}
Separating the input image into transmission and reflection components is an under-constrained problem. Especially when reconstructing a scene using the 3D Gaussians, due to its high flexibility, it tends to overfit virtual reflections behind mirrors or glass using numerous low-opacity Gaussians, leading to ambiguous geometry that becomes entangled with real geometry located behind the reflective surface, as shown in Fig.~\ref{fig:gsfloater}. Based on this observation, we propose the depth variance loss, encouraging Gaussians to form distinct surface structures. We first render a depth map $\mathbf{D}_t$ using Equ.~\ref{equ:depth_map} with primary Gaussians. Then we compute the depth variance loss $\mathcal{L}_d$ as follows:
\begin{equation}
    \label{eqn:ct}
    \mathcal{L}_d=\mathbf{H} \sum_{i=1}\| z_i-\mathbf{D}_t(\mathbf{u})\|_2 o_iG_i(\mathbf{u})T_i,
\end{equation}
where $z_i$ is the depth of the Gaussian center and $\mathbf{H}$ is the hit mask described in Section~\ref{sec:reflection_strength}. Multiplying by $\mathbf{H}$ avoids degrading the quality of non-reflective regions. By minimizing $\mathcal{L}_d$, the primary Gaussians are encouraged to be close to the predicted surfaces, and the reflection components are decomposed from the transmission components.

\FigGSFloater

\subsubsection{Gradient conflict loss}
Drawing inspiration from prior reflection separation works~\cite{levin2007user, xue2015computational,gao2024planar}, we apply the gradient conflict regulation $\mathcal{L}_c$. It is based on the observation that significant color gradients that appear in the reflection image are unlikely to appear in the transmission image:
\begin{equation}
    \label{eqn:gc}
    \mathcal{L}_c=\mathbf{H} \| \nabla(\mathbf{C_t}) \cdot
    sg(\nabla(\mathbf{C_r})) \|_1 \:,
\end{equation}
where $\nabla$ is the Sobel operator. We compute the color gradients of $\mathbf{C}_t$ and $\mathbf{C}_r$, then minimize their dot product, which selectively removes the reflection components in the transmission image. And $sg(\cdot)$ means that we stop the gradient flow from $\mathcal{L}_c$ to $\nabla(\mathbf{C_r})$, ensuring that the reflection image is not corrupted. 

\subsubsection{Reflection mask loss}
To constrain the reflection mask $\mathbf{M}(\mathbf{u})=1$ only at mirror or glass regions, we apply an L1 supervision using manually annotated reflection masks $\mathbf{M}_{a}$ (detailed in next section):
\begin{equation}
    \mathcal{L}_m=\| \mathbf{M}(\mathbf{u}) - \mathbf{M}_{a} \|_1,
\end{equation}
which enforces the reflection regions to stay within the annotated areas while still allowing a learning-based refinement.




\subsection{Training}
\label{sec:training}

\subsubsection{Reflection plane initialization}
\label{sec:train_init}
Recall that the reflection image is rendered by mirroring the primary Gaussians across to the reflection plane. A randomly initialized reflection plane leads to meaningless reflection images, making optimization infeasible. Therefore, we propose a method to initialize the reflection plane using a small number of manually annotated reflection masks.

Specifically, we annotate 5-10 reflection masks $\mathbf{M}_{a}$ with Segment Anything (SAM)~\cite{kirillov2023segment}. For scenes with multiple glass panes, we annotate them with distinct class labels. Then the parameters of reflection planes can be estimated by masks and their corresponding camera poses (see supplementary materials for details). Due to the difficulty of initializing point clouds on reflective surfaces via SfM, we uniformly sample 1000 points on each reflective plane to ensure that the initial Gaussians cover the entire reflective area. We initialize these Gaussians with $m_i=1$, while the Gaussians initialized from SfM are set to 0.



\subsubsection{Loss function}
The overall loss function consists of five terms:

\begin{equation}
    \begin{aligned}
    \mathcal{L} = & (1 - \lambda) \mathcal{L}_1 + \lambda \mathcal{L}_{D-SSIM} \\
                 & + \lambda_{d} \mathcal{L}_d+ \lambda_{c} \mathcal{L}_c+ \lambda_{m} \mathcal{L}_m,
    \end{aligned}
\label{equ:full_loss}
\end{equation}
where $\mathcal{L}_1$ and $\mathcal{L}_{D-SSIM}$ are the same image losses as in 3DGS~\cite{3dgs}. We set $\lambda =0.2$, $\lambda_d = 0.005$, $\lambda_c = 0.2$ and $\lambda_m = 0.5$ for all scenes.

\subsubsection{Opacity perturbation}
We observe that, even with the aforementioned regulations, the transmission image may still retain some residual reflection components. To avoid getting stuck in such local optima, we choose to periodically add noise to the opacity of the primary Gaussians which are behind the reflection plane.  

Firstly, under all training viewpoints, we project the primary Gaussians onto the reflection plane. Gaussians with the projections of their centers falling inside the plane are selected. Then we add a uniformly distributed noise in the range of $[-0.4, 0.4]$ to the opacities of the selected Gaussians and clamp the results every 1000 iterations. To avoid conflict with the opacity clamping operation in the vanilla 3DGS, we interleave opacity-perturbation periods with opacity-clamping periods. 


\subsubsection{Multi-stage optimization}
\label{sec:progressive_opt}
The optimization is divided into three stages to help transmission and reflection decomposition.

\textbf{Stage 1: initialization}. In the first stage, we directly adopt the transmission image as the final output $\mathbf{C} = \mathbf{C}_t$ and optimize the primary Gaussians via $\mathcal{L}_1$ and $\mathcal{L}_{D-SSIM}$. 
\begin{equation}
    \mathcal{L} = (1 - \lambda) \mathcal{L}_1 + \lambda \mathcal{L}_{D-SSIM}. 
\end{equation}

This stage follows the same pipeline as vanilla 3DGS and runs for 3,000 iterations. Also, we initialize the reflection plane $\mathbf{P}$ as described in Section~\ref{sec:train_init} at this stage.

\textbf{Stage 2: reflection plane adjustment}.
In the previous stage, both the reflection and transmission components were fitted by the primary Gaussians. Therefore, in this stage, we leverage the reflection image as a reference to remove ``reflection ghost'' from the transmission, and guide the refinement of the reflection plane. To achieve this, we enable the full rendering equation (Equ.~\ref{equ:rendering_equ}) and loss functions (Equ.~\ref{equ:full_loss}), but block the gradients from the reflection image back to the Gaussians, avoiding coupling between the optimization of the reflection image and the reflection plane. This stage also lasts for 3,000 iterations.


\textbf{Stage 3: joint optimization}.
In the previous stage, we have obtained a reasonably accurate reflection plane. In this stage, we reduce the learning rates of the plane's position and normal (all multiplied by 0.1) to further fine-tune it, and allow full gradient propagation. Opacity perturbation is also enabled at this stage. We allocate 24,000 iterations to stage 3, maintaining the total 30,000 iterations consistent with the vanilla 3DGS.

\newif\ifincludeimages
\includeimagestrue 

\section{Experiments}

\subsection{Implementation Details}
Our method is implemented based on the Pytorch framework of vanilla 3DGS. The rendering process consists of two main passes. Given a camera pose, the first pass takes the primary Gaussians, glass Gaussians and the reflection plane as input and renders both $C_t$ and $R$. The second pass takes the mirrored Gaussians as input to render $C_r$. The final result is obtained by composing these outputs. During training, we adopt the same densification strategy in 3DGS. To comprehensively validate the rendering quality improvements brought by our representation for scenes with glass or mirrors, we integrate 3DGS-MCMC~\cite{3dgs-mcmc} as an enhanced baseline for comparison. All experiments are conducted on a single NVIDIA RTX 4090 GPU.

\subsection{Datasets}
\label{sec:dataset}
We use a self-captured dataset and a publicly available dataset for comparison.

\textbf{Real transmission-reflection dataset (RTR)}. To the best of our knowledge, there is no existing public dataset featuring 360-degree scenes with both transmission and reflection through glass panes. Therefore, we captured six real-world scenes that include common glass windows and showcases. Five of these scenes contain a single glass pane, while one features multiple panes. Each scene consists of 100–200 images captured at a resolution of 960×720.

\textbf{Mirror-NeRF dataset}. We use the real-world dataset introduced by Mirror-NeRF~\cite{mirror_nerf} to evaluate performance under mirror reflection scenarios. This dataset includes three scenes with mirrors, each comprising 260–320 images at a resolution of 960×720.

\subsection{Baselines and Metrics}
\label{sec:baselinemetric}
On the RTR dataset, we compare our method and our MCMC-enhanced method with the following baselines: 3DGS~\cite{3dgs}: vanilla 3D Gaussian Splatting without any reflection modeling; Spec-Gaussian~\cite{spec-gaussian}: a 3DGS-based method which utilizes an ASG appearance field to model specular and anisotropic components; EnvGS~\cite{xie2024envgs}: a 3DGS-based method which renders the reflection color with a ray-tracing-based renderer; Ref-NeRF~\cite{ref-nerf}: a reflection-specialized NeRF-based method; NeRFReN~\cite{nerfren}: a method employing two separate neural radiance fields for transmission and reflections (originally designed for forward-facing scenes); MS-NeRF~\cite{ms-nerf}: a NeRF-based method that decomposes the scene into parallel feature subspaces; 3DGS-MCMC~\cite{3dgs-mcmc}: a 3DGS-based method that modifies the densification and pruning strategy as a deterministic state transition of MCMC samples to enhance rendering quality. On the mirror reflection dataset, we additionally include Mirror-NeRF~\cite{mirror_nerf}: a NeRF-based method that models reflections using Whitted-style ray tracing.


For quantitative evaluations, we select Peak Signal-to-Noise Ratio (PSNR), Structural Similarity Index Measure (SSIM), and Learned Perceptual Image Patch Similarity (LPIPS)~\cite{lpips} as metrics. On the RTR dataset, all methods were evaluated at the same resolution of 960×720 for fair comparison. On the mirror reflection dataset, Mirror-NeRF struggles to converge at high resolutions, so we keep its original setting with a resolution of 480×360. Other methods are evaluated at a resolution of 960×720.

\subsection{Comparison}
\label{sec:comparison}
We conducted comparisons in terms of rendering quality, decomposition result, and efficiency.

\FigRqOurs
\TableRQours
\FigRqmirrornerf

\textbf{Rendering quality}. 
As quantitatively demonstrated in Table~\ref{table:rqours} and Table~\ref{table:rqmirrornerf}, our method achieves higher quality compared to the baseline approaches. And after integrating 3DGS-MCMC, our method achieves a further improvement compared to our 3DGS-based implementation and the 3DGS-MCMC baseline.
The visual comparisons on the RTR dataset are shown in Fig.~\ref{fig:rqours}. 
NeRFReN and MS-NeRF produce blurry reflections due to ineffective transmission–reflection decomposition in general scenes. 3DGS exhibits a noisy mixture of transmission and reflection under novel views as seen in \textit{Ship} and \textit{Bookcase2}. Spec-Gaussian and EnvGS produce almost no reflections as shown in \textit{Gundam} and \textit{Bookcase2} because neither method can accurately model transparent glass panes from SfM initialized point clouds. In contrast, our approach, which integrates 3DGS with the explicitly defined reflection plane and the physically based reflection model, produces high-quality reflections on transparent glass surfaces. Fig.~\ref{fig:rqmirror} also shows visual comparisons on the Mirror-NeRF dataset, which focuses solely on mirror reflections. The comparison shows that our method generalizes well to pure mirror reflections and achieves competitive results compared to other approaches.

\FigDeours
\TableMirror


\FigDeccompare

\textbf{Decomposition visualization}.
The enhanced rendering quality of our method stems from its effective transmission-reflection decomposition. Fig.~\ref{fig:decomours} illustrates our decomposition results, where the two components are accurately separated and recombined to produce high-quality novel view images.
We also compare the decomposition results with NeRFReN~\cite{nerfren}, MS-NeRF\cite{ms-nerf}, and EnvGS~\cite{xie2024envgs}.
As shown in Fig.~\ref{fig:deccompare}, 
NeRFReN degenerates into a single radiance field because its geometric prior assumes that the reflected scene has a simple shell-like geometry, which cannot generalize to wide viewing angles.
The transmission-reflection separation in MS-NeRF is not clean, with residual components leaking into each other. This is because it directly fits the image using multiple subspace radiance fields without applying any decomposition constraints. EnvGS suffers from poor reflections as it only supports tracing rays on opaque surfaces and ignores view-dependent Fresnel effects. By comparison, our method accurately decomposes transmission and reflection on glass panes.


\TableSpeed

\textbf{Efficiency}. 
Table~\ref{table:speed} lists the training time, rendering speed, and number of Gaussians on RTR dataset.
NeRF-based methods require hours of training and also suffer from slow rendering speeds. Among the compared methods, 3DGS serves as a baseline in terms of performance. Spec-Gaussian needs to obtain the view-dependent color of each Gaussian through an MLP, which reduces FPS from 338 to 181. EnvGS performs ray-traced reflections, leading to slow rendering (13 FPS) and long training times (2 hours). Our method has a moderate training time and achieves a rendering speed approximately two-thirds that of 3DGS. In terms of the number of Gaussians, 3DGS models reflections by adding extra Gaussians behind the reflection plane, leading to a larger number of Gaussians compared to our approach. EnvGS introduces an extra set of environment Gaussians to model reflections, significantly increasing the total number of primitives.


\TableAblationRegulation
\FigAbRegulation

\subsection{Ablation Study}
\label{sec:ablation}



We conducted a series of ablation studies to validate the effectiveness of several designs of our method and demonstrate their importance. The experiments are conducted on the \textit{Ship} and \textit{Bookcase2} scenes. The quantitative results of ablation studies are reported in Table~\ref{table:ablation}.

\FigAbReflInsensity
\FigAbSchlick

\textbf{Regulations and opacity perturbation}. 
We conduct ablation studies to evaluate the contributions of the regularization terms and opacity perturbation under various settings: without 
depth variance loss $\mathcal{L}_d$ and gradient conflict loss $\mathcal{L}_c$, with only one of the regulations, and without opacity perturbation. The qualitative ablation results are shown in Fig.~\ref{fig:ablationregulation}.
When no regulations are applied (row 6), the transmission image contains many ``reflection ghosts'', leading to a noisy output image. Removing the depth variance loss $\mathcal{L}_d$ (row 5) leads to poorly constrained Gaussian geometry, causing some Gaussians to incorrectly model the reflected LED tubes which are far from the actual surface of the bookcase. Without the gradient conflict loss $\mathcal{L}_c$ (row 4), the transmission image contains needle-like floaters in areas where the gradient resembles that of the reflection image. Without opacity perturbation (row 3), the decomposition remains incomplete, with residual reflections still embedded in the transmission image, which also degrades the quality of novel view synthesis. These results demonstrate that both the proposed regulations and opacity perturbation are crucial for achieving robust decomposition and high-quality image synthesis.

\textbf{Reflection intensity and Fresnel-based reflection}.
This ablation study is designed to validate the necessity of reflection intensity maps and Fresnel-based view-dependent reflection effects. For the former, we remove the reflection intensity map by setting $\mathbf{C}_r = \mathbf{C}_{raw}$. For the latter, we replace the Schlick reflection model with a view-independent learnable reflection coefficient $R=R_0$. The visual results are shown in Fig.~\ref{fig:ablationreflection_intensity} and Fig.~\ref{fig:ablationschlick}. 
Without reflection intensity (row 3 in Fig~\ref{fig:ablationreflection_intensity}), the reflection image fed into our model no longer conforms to the required linear color space, making it difficult to model accurate reflections. 
Without Fresnel-based reflection (row 3 in Fig.~\ref{fig:ablationschlick}), the model fails to capture the view-dependent variation in reflection strength, especially at grazing angles where reflections become more pronounced, resulting in blurrier and less realistic results. Both lead to a degradation in the final image quality.

\textbf{Mutli-stage optimization}. To evaluate the contribution of our multi-stage optimization, we bypass stage 2 and launch joint optimization immediately after stage 1. As Fig.~\ref{fig:ablationstage} illustrates, this ablation causes the position and orientation of the reflection plane to drift markedly from the true glass geometry. The misalignment prevents the reflection image from fitting the reflection components in the input images, resulting in poor quality due to the failure to separate reflection from transmission.
\FigAbMultiStage

\subsection{Limitation and Future Work}
Although our method has achieved significant progress in modeling transmission and reflection on glass panes, it has two major limitations: handling curved surfaces and multiple non-parallel planes. Our approach relies on the symmetry of planar reflection, making it inapplicable to curved surfaces. Additionally, scenes containing multiple reflection planes require iterative rendering of reflection colors for each plane, which increases training and rendering times. Future work could focus on optimizing the rendering pipeline and parallelizing rendering and optimization across multiple planes to improve efficiency.

\section{Conclusion}
In this paper, we present TR-Gaussians to enhance 3DGS in rendering scenes with transparent glass panes. By explicitly modeling both transmission and reflection through 3D Gaussians and learnable reflection planes, our method effectively captures the complex appearance of transparent glass panes with high fidelity. A multi-stage optimization framework, combined with tailored constraints and an opacity perturbation strategy, enables accurate decomposition and high-quality rendering. Extensive experiments on different datasets validate the effectiveness and efficiency of our approach.



 
\bibliographystyle{IEEEtran}
\bibliography{texts/egbib}

\onecolumn
\newpage
\twocolumn

\appendices

\section{Details of estimating plane parameters}
We estimate the reflection plane from the annotated masks as follows:

\begin{enumerate}
    \item Glass boundary extraction. For each annotated mask $\mathbf{M}_a$, we extract its edge and dilate it by 20 pixels. The resulting expanded region is defined as the glass boundary to filter candidate points.

    \item Candidate selection via projection. We project the primary Gaussians onto the views with annotated mask. Only those Gaussians whose centers fall within the glass boundary are selected as candidates for plane fitting.

    \item Plane fitting with RANSAC. Using the selected Gaussians, we fit the plane via RANSAC and get the plane normal $\mathbf{n}_p$. The centroid of the inlier positions found by RANSAC are then computed to define the plane center $\mathbf{u}_p$.

    \item Extent determination via 2D bounding box. We project the inliers onto the estimated plane and obtain a set of 2D points. We then apply the rotating calipers algorithm to compute the minimum-area oriented bounding box (OBB). The width $w$ and height $h$ of this box define the spatial extent of the reflection plane.

    \item Plane initialization. Finally, we initialize the reflection plane as $\mathbf{P} = (\mathbf{u}_p, \mathbf{n}_p, w, h, R_0)$. The initial base reflectance $R_0$ is set depending on the material: $R_0=0.2$ for transparent glass, and $R_0=1.0$ for mirrors.
\end{enumerate}

\section{Rendering multiple reflection planes}
Our method is capable of extending to scenes with $N_p \geq 2$ reflection planes by iterative rendering. Specifically, we compute the reflection strength image $\mathbf{M}_i$, hit mask $\mathbf{H}_i$ for each plane, and reflect primary Gaussians about each plane to render reflection images $\mathbf{C}_r^i$. Then we aggregate per-plane reflection strength image $\mathbf{R}_i$ and reflection image $\mathbf{C}_i^r$  based on hit mask $\mathbf{H}_i$ to produce the composite outputs: 
\begin{equation}
    \mathbf{R}=\sum_i^{N_p} \mathbf{H}_i  \mathbf{R}_i, \:
    \mathbf{C}_r=\sum_i^{N_p} \mathbf{H}_i  \mathbf{C}_r^i.
\end{equation}

\section{Extension with 3DGS-MCMC}
We integrate 3DGS-MCMC~\cite{3dgs-mcmc} as an enhanced baseline into our framework to further improve rendering quality. Specifically, we substitute the densification process in vanilla 3DGS~\cite{3dgs} with the relocation strategy from 3DGS-MCMC~\cite{3dgs-mcmc} and integrate its opacity and covariance regulation into our loss function: 
\begin{equation}
    \begin{aligned}
    \mathcal{L} = & (1 - \lambda) \mathcal{L}_1 + \lambda \mathcal{L}_{D-SSIM} \\
                 & + \lambda_{d} \mathcal{L}_d+ \lambda_{c} \mathcal{L}_c+ \lambda_{m} \mathcal{L}_m \\
                 & + \lambda_o\mathcal{L}_o + \lambda_{\sum}\mathcal{L}_{\sum},
    \end{aligned}
\end{equation}
where we set $\lambda_o=0.01$ and $\lambda_{\sum}=0.01$, following the original paper. 
3DGS-MCMC also requires setting a maximum capacity  for the number of Gaussians. For fair comparison, we set it to match the number of Gaussians obtained by our original method. 



\section{Decomposition results on Mirror-NeRF dataset}
\begin{figure}[h]
  \begin{center}
    \includegraphics[width=0.49\textwidth]{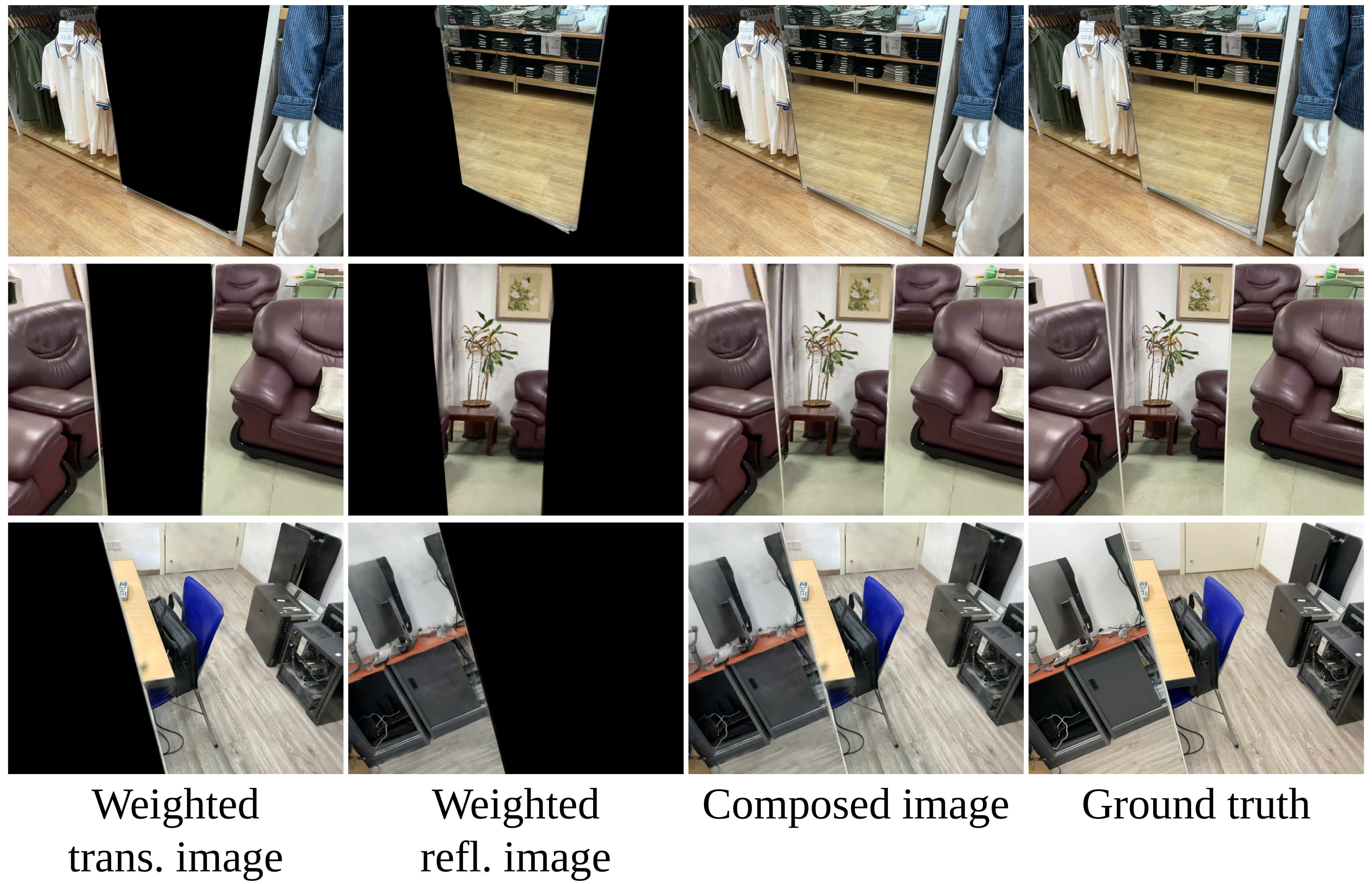}
  \end{center}
  \vspace{-5mm}
  \caption{Decomposition results on Mirror-NeRF dataset. From top to bottom: Market, Lounge, and Discussion room. From left to right: weighted transmission image $\mathbf{C}_t\cdot(1-\mathbf{R})$, weighted reflection image $\mathbf{C}_r\cdot\mathbf{R}$, composed image $\mathbf{C}$, and ground truth.}
  \label{fig:decommirror}
\end{figure}

We further present qualitative decomposition results on the Mirror-NeRF~\cite{mirror_nerf} dataset, as shown in Fig.~\ref{fig:decommirror}. Our method accurately separates reflection components in scenes with mirrors. This demonstrates that our approach is not limited to transparent glass panes, but also generalizes well to purely specular reflections. This capability stems from our Fresnel-based reflection model, which explicitly accounts for both transmission and reflection phenomena in a unified framework.

\section{Ablation study on integration with 3DGS-MCMC}

\begin{figure}[h]
  \begin{center}
    \includegraphics[width=0.49\textwidth]{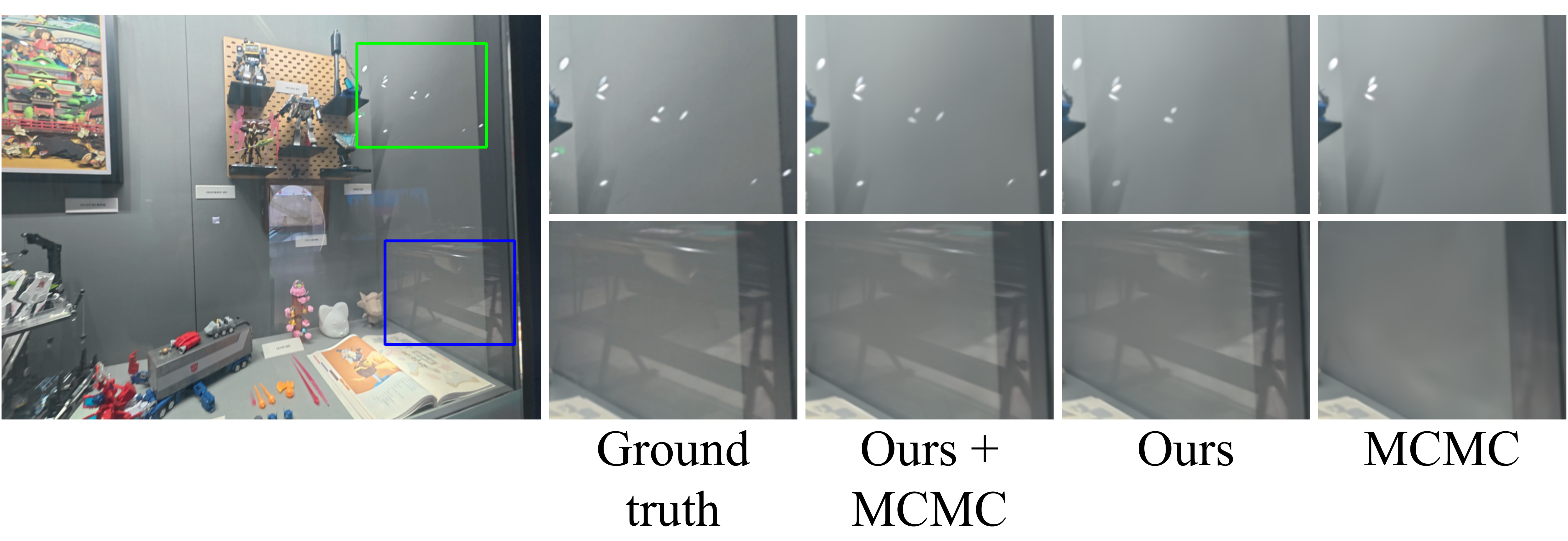}
  \end{center}
  \vspace{-5mm}
  \caption{From left to right: ground truth, our method integrated with 3DGS-MCMC, our method and 3DGS-MCMC.}
  \label{fig:ablationmcmc}
\end{figure}

To validate the generalizability of our representation, we integrate the 3DGS-MCMC~\cite{3dgs-mcmc} baseline into our TR-Gaussians pipeline to further enhance image quality. The qualitative results are shown in Fig.~\ref{fig:ablationmcmc}.
As 3DGS-MCMC does not explicitly model reflections, it exhibits similar limitations to the vanilla 3DGS: reflection components are either missing in novel viewpoints (first row) or appear blurred (second row). In contrast, our method incorporates an explicitly defined reflection plane and a Fresnel-based reflection model into the 3DGS framework, enabling the synthesis of high-fidelity reflections on transparent glass surfaces. Furthermore, by adopting the densification and pruning strategy from 3DGS-MCMC, our method can capture finer reflection details, such as the complete reconstruction of multiple reflected light bulbs (first row), which are only partially recovered by other baselines. These results demonstrate that our representation can be seamlessly integrated into advanced 3DGS variants like 3DGS-MCMC, leading to enhanced reconstruction quality and more accurate reflection modeling.


\end{document}